%Paper: hep-th/9302077
%From: dij@s-a.amtp.liv.ac.uk
%Date: Wed, 17 Feb 1993 16:44:03 +0000 (GMT)
%Date (revised): Thu, 6 May 1993 16:06:30 +0100 (BST)

\def \s{\sigma}
\def \b{\beta}
\def \a{\alpha}
\def \g{\gamma}
\def \d{\delta}

\def \l{\lambda}

\def \ph{\phi}

\def \Ph{\Phi}
\def \p{\pi}

\def \D{\Delta}

\def \m{\mu}
\def \n{\nu}
\def \ph{\phi}

\def \O{\Omega}
\def \D{\Delta}
\def \r{\rho}

\def\apny{Ann.\ Phys.\ (New York)\ }
\def\cmp{Comm.\ Math.\ Phys.\ }

\def\ijmpa{{Int.\ J.\ Mod.\ Phys.\ }{\bf A}}

\def\mpla{{Mod.\ Phys.\ Lett.\ }{\bf A}}
\def\nc{Nuovo Cimento\ }
\def\npb{{Nucl.\ Phys.\ }{\bf B}}

\def\plb{{Phys.\ Lett.\ }{\bf B}}

\def\pr{Phys.\ Rev.\ }
\def\prd{{Phys.\ Rev.\ }{\bf D}}
\def\prl{Phys.\ Rev.\ Lett.\ }

\def\lmp{Lett.\ Math.\ Phys.\ }

\input harvmac
\def \del {\partial}
\def \bdel{\bar \del}
\def \bz {\bar z}
\def \cG {{\cal G}}
\def \bJ{\bar J}
\def \bA{\bar A}
\def \cJ{{\cal J}}
\def \ad{{\rm ad}}
\def \cC{{\cal C}}
\def \rank{{\rm rank}}
\def \bs{\setminus}
\def \wr{\tilde r}
\Title{LTH 304}{WZW-Toda Reduction using the Casimir Operator}
\centerline{I. Jack and J. Panvel}
\bigskip
\centerline{\it DAMTP, University of Liverpool, Liverpool L69 3BX, U.K.}
\vskip .3in
We construct a quantum Hamiltonian operator for the Wess-Zumino-Witten (WZW)
model
in terms of the Casimir operator. This facilitates the discussion
 of the reduction of the WZW model to Toda field theory at the quantum level
and provides a very straightforward derivation of the quantum central charge
for the Toda field theory.
\Date{February 1993}

\newsec{Introduction}
Two-dimensional
conformal field theories have been the object of much attention, in relation
 to string theory and also to statistical mechanics. The Wess-Zumino-Witten
(WZW)
models\ref\ew{E. Witten, \cmp92 (1984) 483.}  based on a Lie group $\cG$
play a central role in the study and classification of conformal field
theories, since a large variety  of rational conformal field theories can be
obtained by gauging the WZW model. On the one hand, the Goddard-Kent-Olive
 (GKO) construction\ref\GKO{P. Goddard, A. Kent and D. Olive, \plb152 (1985)
88.}
 (which, for instance, generates the minimal models) can be
realised by gauging a diagonal subgroup of $\cG$\ref\wzwg{K. Gawedzki and
A. Kupiainen, \plb215b (1988) 119;\npb320 (1989) 625\semi
D. Karabali and H. J. Schnitzer, \npb329 (1990) 649.}.
 On the other hand,
  Toda field theories\ref\toda{A. N. Leznov and M. V. Saveliev,
\lmp3 (1979) 489; \cmp74 (1980) 111.}
 can be obtained by gauging a nilpotent subgroup of a maximally non-compact
group
 $\cG$\ref\dubI{J. Balog, L. Feh\'er, L. O'Raifeartaigh, P. Forg\'acs and
A. Wipf, \apny203 (1990) 76; \plb244 (1990) 435.}.
 In the Toda case, the function of the gauging is essentially to
impose constraints\ref\dubII{P. Forg\'acs, A. Wipf, J. Balog, L. Feh\'er
and L. O'Raifeartaigh, \plb227 (1989) 214.}
 on the Kac-Moody currents associated with the WZW model.
The discussion of the GKO construction and the reduction of the WZW model to
Toda theory can both be facilitated by the recent discovery of a free-field
representation of the WZW model\ref\mosc{A. Gerasimov, A. Marshakov,
A. Morozov, M. Olshanetsky and S. Shatashvili, \ijmpa5 (1990) 2495.}
\ref\suz{M. Kuwahara and H. Suzuki, \plb235 (1990) 52, 57.}.
 This was found useful in the discussion of
 the BRST quantisation of the gauged WZW models corresponding to Toda field
 theory \ref\hay{N.
 Hayashi, \npb363 (1991) 681.}.

In particular, the simple form for the energy-momentum tensor of the
WZW model in the free-field representation leads straightforwardly to the
correct expression for the central charge of the Toda field theory at the
quantum level\ref\man{P. Mansfield, \npb208 (1982) 277; \npb222 (1983) 419.}.
Our purpose here is to discuss the derivation of the Toda field theory from
the WZW model in a way which parallels the free-field approach, but which
 is more rooted in group theory than in conformal field theory. The method is
derived from techniques used recently\ref\dvv{R. Dijkgraaf, E. Verlinde
and H. Verlinde, \npb371 (1992) 269.}
 to derive the exact metric and dilaton
fields for string black holes (which are also obtained by gauging a WZW
 model)\ref\sbh{E. Witten, \prd44 (1991) 314.}.
 The idea is to construct the Hamiltonian operator for the WZW model,
incorporating quantum effects using the Sugawara construction\ref\sug
{H. Sugawara, \pr170 (1968) 1659; for a review see also
P. Goddard and D. Olive, \ijmpa1 (1986) 303.},
 and also
including additional terms arising from the gauging.
(This procedure is similar to results that have emerged in the discussion\ref
\Mor{A. Yu. Morozov, A. M. Perelomov, A. A. Rosly, M. A. Shifman and A. V.
Turbiner, \ijmpa5 (1990) 803.}
of the analogy between the Hamiltonians of quasi-exactly-soluble quantum
mechanical systems\ref\tur{A. V. Turbiner, \cmp118 (1988) 467; M. A. Shifman
and A. V. Turbiner, \cmp126 (1989) 347; M. A. Shifman, \ijmpa4 (1989) 2897.}
 and the energy-momentum tensor of conformal theories
based on the generalised Sugawara construction\ref\halp{M. B. Halpern and
E. B. Kiritsis,\mpla4 (1989) 1373;\mpla4 (1989) 1797(E).}.)
 The field equation
 derived from this Hamiltonian is then compared with the field equation
 deriving from the string effective action for a general non-linear
 $\s$-model, from which the metric and dilaton fields can be read off, and
thence the action for the Toda field theory can be reconstructed. In the case
of Witten's string black hole\sbh\ the method provided an exact solution
whose validity has been checked up to fourth order in perturbation theory
\ref\jjp{A. A. Tseytlin, \plb268 (1991) 175; I. Jack, D. R. T. Jones and
J. Panvel, \npb393 (1993) 95.}.
 The method has also been used to construct
more general exact string black hole solutions\ref\bars{I. Bars and K.
Sfetsos, \prd46 (1992) 4495, 4510; ``$SL(2,R)\times SU(2)/R^2$ string model
in curved spacetime and exact conformal results'', preprint USC-92/HEP-B3.}.
Similar results have recently been obtained by considering the quantum
effective action for the gauged WZW model\ref\aatI{A. A. Tseytlin,
``Effective action of gauged WZW model and exact string solutions'',
preprint Imperial/TP/92-93/10 (hep-th/9301015); I. Bars and K. Sfetsos,
``Exact effective action and spacetime geometry in gauged WZW models'',
preprint USC-93/HEP-B1 (hep-th/9301047).}.
The Hamiltonian operator for the WZW model is (up to a factor) precisely the
quadratic Casimir operator for the group $\cG$, the zero-modes of the
Kac-Moody currents being the left- and right-invariant vector fields on the
group manifold. A useful co-ordinate system on the group manifold for the
present purpose is that corresponding to the Gauss decomposition of the group
elements for a maximally non-compact group $\cG$.
In this co-ordinate system, the quadratic Casimir operator assumes
a particularly simple form, which can be given explicitly for a general
maximally non-compact Lie
group\ref\ls{A. N. Leznov and M. V. Saveliev, Soviet J. Part. Nucl. {\bf 7}
(1976) 22.}\nref\fls{I. A. Fedoseev, A. N. Leznov and M. V. Saveliev,
\nc A76 (1983) 596.}--\ref\book{A. N. Leznov and M. V. Saveliev,
Group-theoretical methods for integration of non-linear dynamical
systems (Birkh\"auser, Basel, 1992).}.
 This simple form is
analogous (though not exactly equivalent) to the free-field representation
 of the WZW Lagrangian. The correct quantum conformally-invariant action for
 the Toda field theory then follows almost immediately. We feel that this
derivation of the Toda field theory from the WZW model is quicker and more
explicit than the use of the free-field representation. Moreover, the method
 may be useful for discussing the so-called ``non-abelian''
Toda theories\ref\natI{A. N. Leznov and M. V. Saveliev, \lmp6
(1982) 505; \cmp89 (1983) 59.}\ref\gervais{J.-L. Gervais and M. V.
Saveliev, \plb286 (1992) 271.}
based on a non-canonical grading of the Lie algebra of $\cG$, which may be
related to string black holes. Although the central charge of these theories
has been calculated\ref\dubIII{L. O'Raifeartaigh and A. Wipf,
\plb251 (1990) 361.},
 the full quantum action does not appear to be known.

\newsec{The Wess-Zumino-Witten model}
The action for the Wess-Zumino-Witten model defined on a group manifold $\cG$
 is given by\ew\
$$
kS(g)=-{k\over{8\p}}\int d^2x \tr(\del_{\m} g\del^{\m}g^{-1})
+{k\over{12\p}}\int \tr(dg\,g^{-1})^3  \eqno(2.1)
$$
where $g\in\cG$. (We assume for convenience that the trace is normalised
 to agree with the trace in the adjoint representation.)
This action is invariant under the transformations
$$
g(z,\bz)\rightarrow\O_L(z)g(z,\bz)\O_R(\bz) \eqno(2.2)
$$
whose generators are the currents
$$
 J(z)=k\del gg^{-1},\qquad \bJ(\bz)=-kg^{-1}\bdel g \eqno(2.3)
$$
which generate two commuting copies of the Kac-Moody algebra.
We shall consider a maximally non-compact Lie group $\cG$ and
make use of the Gauss decomposition of the group element $g\in\cG$, writing
$$
g=g_-g_0g_+ \eqno(2.4)
$$
where
$$
g_-=\exp(\sum_{\a\in\Ph+}\ph^{\a}_-E_{-\a}),\qquad
 g_+=\exp(\sum_{\a\in\Ph+}\ph^{\a}_+E_{\a}),
\qquad g_0=\exp(\sum_{i=1}^{\rank\cG} r^iH_i).  \eqno(2.5)
$$
In Eq. (2.5), $\Ph^+$ is the set of positive roots $\a$
 and $E_{\a}$ are the
corresponding step operators, while $\{H_i,i=1,2,\ldots \rank\cG \}$ are the
generators of the Cartan subalgebra. (See the Appendix for our Lie algebra
conventions.)
The parameters $\{\ph_-^{\a},\ph_+^{\a},r^i\}$ may be
regarded as co-ordinates on the group manifold $\cG$.

As mentioned in Section 1, the Toda field theory is obtained from the WZW
 model by gauging in order to impose certain constraints\dubI\dubII.
 In the usual
gauging of the WZW model\wzwg, one gauges the vector invariance
$g\rightarrow\g g\g^{-1}$, $\g(z,\bz)\in H$ for some subgroup $H$ of $\cG$.
However, owing to the nilpotency of the groups $\cG_{\pm}$ generated by
elements
of the form $g_{\pm}$ in Eq. (2.5), in this case we may gauge the invariance
$g\rightarrow\a g\b^{-1},\quad \a(z,\bz)\in\cG_-,\quad \b(z,\bz)\in\cG_+$.
The gauged action is\dubI
$$
kS_G(g,A,\bA)=kS(g)+{k\over{4\p}}\int d^2x \tr(\bA\del gg^{-1}+Ag^{-1}\bdel g
+\bA gAg^{-1}-\bA\m-A\n). \eqno(2.6)
$$
 The conventional vector gauging sets to zero the currents corresponding to the
 subgroup $H$ of $\cG$. Here, however, the presence of the matrices $\m$, $\n$
defined by
$$
\m=\sum_{\a\in\D}\m^{-\a}E_{-\a}, \qquad
\n=\sum_{\a\in\D}\n^{\a}E_{\a}   \eqno(2.7)
$$
(where $\D$ denotes the set of simple roots)
means that the effect of the gauging is to impose the constraints
$$\eqalignno{
J_{\a}\equiv\tr(E_{\a}J)=k\m_{\a},
\qquad &\bJ_{-\a}\equiv\tr(E_{-\a}\bJ)=-k\n_{-\a} \qquad (\a\in\D) \cr
J_{\a}=0,\qquad &\bJ_{-\a}=0 \qquad (\a\in\Ph^+\bs \D). &(2.8) \cr}
$$
The resultant Toda field theory action is given at the classical level by
$$
S={k\over{8\p}}\int d^2x \bigl(\g_{ij}\del_{\m}r^i\del^{\m}r^j+\sum_{\a\in\D}
M_{\a}e^{-r^i\a(H_i)}\bigr),\eqno(2.9)
$$
where
$$
M_{\a}=\g^{\a(-\a)}\m_{\a}\n_{-\a}\eqno(2.10)
$$
and $\g_{ij}$ is defined in Eq. (A.3).
The energy-momentum tensor for the WZW model is constructed from the
Kac-Moody currents according to the Sugawara construction\sug. However with the
imposition of the constraints Eq. (2.8), the energy-momentum tensor must be
modified by an ``improvement term''\dubI\ to ensure that it commutes with the
constraints. The result is
$$
T(z)={1\over{k-h}}(J,J)-2(\d,\del J) \eqno(2.11)
$$
where $h$ is the dual Coxeter number for the group $\cG$ and where
$$
\d=\sum_{\a\in\Ph^+}{H_{\a}\over{<\a,\a>}} \eqno(2.12)
$$
(with an analogous result for the antiholomorphic component of the
energy-momentum tensor, $\bar T(\bz)$).
The important point to note is that
this form of the energy-momentum tensor fully incorporates quantum effects.
(The unconventional sign of $h$ in Eq. (2.11) arises because we are using the
conventions of Refs. \sbh, \dvv, appropriate for a non-compact group.)
We now introduce vertex operators $V(\ph_-^{\a},\ph_+^{\a},r^i)$ for the
WZW model. The zero modes $J_0$, $\bJ_0$ in Laurent expansions of the
Kac-Moody currents can be expressed as operators $\cJ^L$, $\cJ^R$ acting on
the vertex operators\dvv. The components of these operators, defined by
$$
\cJ^L_a=(T_a,\cJ^L),\qquad \cJ^R_a=(T_a,\cJ^R), \eqno(2.13)
$$
where $T_a$ denotes a generic element of the Lie algebra,
 are in fact the generators of
left and right multiplication by Lie algebra elements, {\it viz.}
$$
\cJ_a^Lg=T_ag, \qquad \cJ_a^Rg=gT_a. \eqno(2.14)
$$
The Virasoro operator $L_0$ can then also be expressed as a
differential operator\dvv\ by replacing the zero
modes of the currents in $T(z)$
in  Eq. (2.11) by $\cJ_L^a$, and similarly $\bar L_0$ may also be obtained
by replacing the zero modes of the currents in $\bar T(\bz)$ by $\cJ_R^a$.
The results are
$$\eqalignno{
L_0&={1\over{k-h}}(\cJ^L,\cJ^L)+2(\d,\cJ^L), \cr
\bar L_0&={1\over{k-h}}(\cJ^R,\cJ^R)+2(\d,\cJ^R). &(2.15) \cr}
$$
The expressions  $(\cJ^L,\cJ^L)$ and $(\cJ^R,\cJ^R)$ are
 in fact both equal to the Casimir operator $\cC$ for the group $\cG$.
 Moreover,
since $\d$ is in the Cartan subalgebra, $(\d,\cJ^L)=(\d,\cJ^R)$.
Hence $L_0$ and $\bar L_0$ coincide as operators. In string theory, the
 physical state conditions for tachyonic vertex operators read
$$
(L_0+\bar L_0-2)V(\ph_-^{\a},\ph_+^{\a},r^i)=0,\qquad
(L_0-\bar L_0)V(\ph_-^{\a},\ph_+^{\a},r^i)=0. \eqno(2.16)
$$
The second of these equations is automatically satisfied; the first becomes,
using Eq. (2.14)
$$
({\cC\over{k-h}}+2(\d,\cJ_R)-1)V(\ph_-^{\a},\ph_+^{\a},r^i)=0. \eqno(2.17)
$$
The operator in Eq. (2.17) may be regarded as the Hamiltonian operator for the
Toda field theory, which already incorporates quantum effects via the
Sugawara construction. We will see later how we can read off from the
Hamiltonian the fully conformally invariant Toda field theory Lagrangian.
Firstly, however, we devote the next Section to a discussion of the form of
 the Casimir operator.

\newsec{The Casimir Operator}

A thorough investigation of the form of the Casimir operator corresponding to
various decompositions of a general Lie algebra $\cG$ has been undertaken by
Leznov and Saveliev\ls. The results are compiled in a recent book\book. For
completeness, however, we will summarise the main points relating to the Gauss
decomposition here, and we will also give more explicit results for some of
 the quantities involved.

As we mentioned in Section 2, the Casimir operator is given by
$$
\cC=(\cJ^L,\cJ^L)=(\cJ^R,\cJ^R) \eqno(3.1)
$$
where $\cJ^L$ and $\cJ^R$ are operators whose components, defined in
Eq. (2.12), are the generators of left and right shifts
respectively. Mathematically they may also be described as left- and
 right-invariant vector fields respectively, defined on the group manifold.
Our first step, following Leznov and Saveliev,
 is to obtain expressions for these shift generators. It is sufficient to
 concentrate on, say, the right shift operators. We consider the component
$\cJ^R_a$ which generates multiplication on the right by $T_a$, where $T_a$
represents one of $E_{\a}$, $E_{-\a}$, or $H_i$. If we wish to be more
specific, we shall denote the operators representing multiplication on the
right by $E_{\a}$, $E_{-\a}$, and $H_i$ as $\cJ_{\a}^R$, $\cJ_{-\a}^R$ and
$\cJ_i^R$ respectively.
 We begin by introducing
 operators $X_{-\a}^L$, $X_{-\a}^R$, $X_{+\a}^L$, $X_{+\a}^R$ defined by
$$\eqalignno{
X_{-\a}^Lg_-=E_{-\a}g_-,&\qquad X_{-\a}^Rg_-=g_-E_{-\a}, \cr
X_{+\a}^Lg_+=E_{\a}g_+, &\qquad X_{+\a}^Rg_+=g_+E_{\a}. &(3.2)\cr}
$$
The operators $X^{L,R}_{-\a}$ only act on $g_-$ and contain only the variables
$\ph^{\a}_-$, and similarly the operators $X^{L,R}_{+\a}$ only act on $g_+$
and contain only the variables $\ph^{\a}_+$. Hence we have
$$\eqalignno{
X^{L,R}_{-\a}g_+=X^{L,R}_{-\a}g_0&=X^{L,R}_{+\a}g_-=X^{L,R}_{+\a}g_0=0,\cr
[X^{L,R}_{-\a},X^{L,R}_{+\a}]&=0. &(3.3)\cr}
$$
It is then clear that
$$
\cJ_{\a}^R=X_{+\a}^R. \eqno(3.4)
$$
In general, though, we have
$$
\cJ_a^Rg=gT_a=g_-g_0(g_+T_ag_+{}^{-1})g_+. \eqno(3.5)
$$
It is convenient to define
$$
E^{\a}=\g^{\a\b}E_{\b},\qquad H^i=\g^{ij}H_j, \eqno(3.6)
$$
where $\g^{ab}$ is the inverse of the Cartan-Killing form $\g_{ab}$,
so that
$$\eqalignno{
(E^{\a},E_{\b})=\d^{\a}{}_{\b},&\qquad (E^{\a},H_i)=0, \cr
(H^i,H_j)=\d^i{}_{j},&\qquad (H^i,E_{\pm\a})=0. &(3.7) \cr}
$$
We can then write
$$\eqalignno{
g_+T_ag_+{}^{-1}&=\sum_{\b\in\Ph^+}((E^{-\b},g_+T_ag_+{}^{-1})E_{-\b}
\cr&\quad
+(E^{\b},g_+T_ag_+{}^{-1})E_{\b})+(H^j,g_+T_ag_+{}^{-1})H_j. &(3.8)\cr}
$$
Substituting Eq. (3.8) into Eq. (3.5), and using
$$
g_0E_{-\b}=e^{-r^i\b(H_i)}E_{-\b}g_0, \eqno(3.9)
$$
followed by Eq. (3.2), we obtain
$$\eqalignno{
\cJ_a^R&=\sum_{\b\in\Ph^+}
((E^{-\b},g_+T_ag_+{}^{-1})e^{-r^i\b(H_i)}X_{-\b}^R
+(E^{\b},g_+T_ag_+{}^{-1})X_{+\b}^L)\cr
&\qquad +(H^j,g_+T_ag_+{}^{-1}){\del\over{\del r^j}}. &(3.10)\cr}
$$
Note that in the case of $\cJ_i^R$, the first term on the right-hand side of
Eq. (3.11) is zero, and in the case of $\cJ_{\a}^R$, the first and
third terms on
 the right-hand side are zero. This expression for $\cJ_{\a}^R$ agrees with
that given earlier in Eq. (3.4) in virtue of the identity
$$
X_{+\a}^R=\sum_{\b\in\Ph^+}(E^{\b},g_+E_{\a}g_+{}^{-1})X_{+\b}^L. \eqno(3.11)
$$
Now
$$
(gLg^{-1},gMg^{-1})=(L,M) \eqno(3.12)
$$
for any elements of the Lie algebra $L$, $M$ and any $g\in\cG$. So from
Eq. (2.13) we have
$$\eqalignno{
\cJ^R&=\sum_{\b\in\Ph^+}
((g_+^{-1}E^{-\b}g_+)e^{-r^i\b(H_i)}X_{-\b}^R+(g_+^{-1}E^{\b}g_+)X_{+\b}^L)\cr
&\qquad + (g_+^{-1}H^jg_+){\del\over{\del r^j}}. &(3.13) \cr}
$$

Using Eqs. (3.2), (3.3), we may also write Eq. (3.13) in the form
$$\eqalignno{
\cJ^R&=\sum_{\b\in\Ph^+}(X_{-\b}^R(g_+{}^{-1}E^{-\b}g_+)e^{-r^i\b(H_i)}
+X_{+\b}^L(g_+{}^{-1}E^{\b}g_+)\cr
&\qquad +g_+{}^{-1}H_{\b}g_+)
+{\del\over{\del r^j}}(g_+{}^{-1}H^jg_+). &(3.14) \cr}
$$
Substituting Eqs. (3.14) and (3.13) for the first and second occurrences
 respectively of $\cJ^R$ on the right-hand side of Eq. (3.1), and using Eq.
(3.12), we obtain finally
$$\eqalignno{
\cC&=\sum_{\b\in\Ph^+}(2\g^{\b(-\b)}e^{-r^i\b(H_i)}X_{-\b}^RX_{+\b}^L
+\g^{ij}\b(H_i){\del\over{\del r^j}})\cr&\qquad
+\g^{ij}{\del\over{\del r^i}}{\del\over{\del r^j}}.
&(3.15) \cr}
$$
In Eqs. (3.10) and (3.15) we have simple expressions for the shift operators
 $\cJ_a^R$ and the Casimir operator $\cC$ in terms of the operators
 $X^R_{-\a}$ and $X^L_{+\a}$. We can in fact obtain fairly explicit
expressions for these operators. It is easy to show that
$$
(\del_{+\a}g_+)g_+{}^{-1}={{e^{\ad\ph_+}-1}\over{\ad\ph_+}}E_{\a} \eqno(3.16)
$$
where
$$
\ph_+=\sum_{\a\in\Ph^+}\ph^{\a}_+E_{\a},\qquad (\ad\ph_+) L=[\ph_+,L]
 \eqno(3.17)
$$
for $L$ in the Lie algebra of $\cG$,
and $\del_{+\a}\equiv{\del\over{\del\ph^{\a}_+}}$. Hence we have
$$
E_{\a}g_+={{\ad\ph_+}\over{e^{\ad\ph_+}-1}}\del_{+\a}g_+. \eqno(3.18)
$$
Using
$$
{x\over{e^{x}-1}}=\sum {{b_n}\over{n!}}x^n, \eqno(3.19)
$$
where $b_n$ are the Bernoulli numbers,
we find, comparing Eqs. (3.2) and (3.18),
$$
X_{+\a}^L=\sum {{b_n}\over{n!}}(\ad \ph_+)^n\del_{+\a}. \eqno(3.20)
$$
By considering the action of $X_{+\a}^L$ on $\ph_+$, we obtain the following
more explicit expression:
$$
X_{+\a}^L=\sum_{n=0}^{\infty}{b_n\over{n!}}
N_{\a\r_1\ldots\r_n}\ph_+^{\r_1}\ldots\ph_+^{\r_n}\del_{+(\a+\r_1+\ldots
+\r_n)} \eqno(3.21)
$$
where $N_{\a\r_1\ldots\r_n}$ is the coefficient of $E_{\a+\r_1+\ldots
\r_n}$ in $[E_{\r_n},\ldots[E_{\r_2},[E_{\r_1},E_{\a}]]\ldots]$.
Similarly, using
$$
g_-{}^{-1}\del_{-\a}g_-=-{{e^{-\ad\ph_-}-1}\over{\ad\ph_-}}E_{-\a} \eqno(3.22)
$$
we can derive
$$
X_{-\a}^R=\sum {{b_n}\over{n!}}(-\ad \ph_-)^n\del_{-\a} \eqno(3.23)
$$
where
$$
\ph_-=\sum_{\a\in\Ph^+}\ph_-^{\a}E_{-\a},\qquad \del_{-\a}\equiv
{\del\over{\del\ph_-^{\a}}}. \eqno(3.24)
$$
Of course $X_{-\a}^L$ and $X_{+\a}^R$ may be obtained by interchanging
the subscripts $+$
and $-$ in Eqs. (3.20), (3.23). In fact, it is easy to see that $X^R_{\a}$
may be obtained from $X^L_{\a}$ by replacing $\ph_+$ by $(-\ph_+)$, and
$X^L_{-\a}$ and $X^R_{-\a}$ may be obtained from $X^L_{\a}$ and $X^R_{\a}$
respectively by
replacing $\ph_+$ by $(-\ph_-)$, and hence explicit expressions may be
obtained for all these operators starting from Eq. (3.21).
 Since the Lie algebras for $\cG_-$ and
$\cG_+$ are nilpotent, the summations in Eqs. (3.20), (3.21) and (3.23)
terminate at the level of the highest root. Combined with the
Baker-Campbell-Hausdorff formula
$$
g_+E_{-\a}g_+^{-1}=\sum{1\over{n!}}(\ad \ph_+)^nE_{-\a} \eqno(3.25)
$$
(which of course also terminates) we may readily obtain
explicit formul\ae for the shift operators in Eq. (3.10) and the Casimir
operator in Eq. (3.15) for any given Lie group. The identity Eq. (3.12)
relating $X_{\a}^L$ and $X_{\a}^R$ is a simple consequence of Eqs. (3.20),
 (3.23), (3.25) and the identity
$$
b_n=\sum_{k=0}^{n}C(n,k)b_k \qquad(n\neq 1). \eqno(3.26)
$$
Finally we note that we can obtain a simple completely explicit formula for
$\cJ_i^R$ as follows. Writing
$$
gH_i=g_-g_0(g_+H_ig_+{}^{-1})g_+, \eqno(3.27)
$$
we have
$$\eqalignno{
 g_+H_ig_+{}^{-1}&=H_i-\sum_{\a\in\Ph^+}\a(H_i)\sum_n{1\over{(n+1)!}}
(\ad\ph_+)^nE_{\a} \cr
&=H_i-\sum_{\a\in\Ph^+}\a(H_i){{e^{\ad\ph_+}-1}\over{\ad\ph_+}}E_{\a}.
&(3.28)\cr}
$$
Hence, using Eq. (3.16), we have
$$
\cJ_i^h={\del\over{\del r^i}}-\sum_{\a\in\Ph^+}\ph_+^{\a}\a(H_i)\del_{+\a}.
 \eqno(3.29)
$$

\newsec{Toda field theory}

Toda field theory is obtained from the Wess-Zumino-Witten model by imposing
the constraints Eq. (2.8). In the context of the vertex operators
$V(\ph_-^{\a},\ph_+^{\a},r^i)$, Eqs. (2.8) are imposed as operator
constraints derived again from the zero modes of the currents. The constraints
Eq. (2.8) imply
$$\eqalignno{
\cJ_{\a}^RV=\m_{\a}V, \qquad& \bar \cJ_{-\a}^LV=-\n_{-\a}V \qquad(\a\in\D)\cr
\cJ_{\a}^RV=0, \qquad& \bar \cJ_{-\a}^LV=0 \qquad(\a\in\Ph^+\bs\D).&(4.1)\cr}
$$
Since $\cJ_{\a}^RV=0$ except for simple roots, we may use Eqs. (3.4), (3.26)
(with $-\a\rightarrow\a$) to obtain
$$
\del_{\a}V=0\quad(\a\in\Ph^+\bs\D),\qquad \del_{\a}V=\m_{\a}V \quad (\a\in\D)
\eqno(4.2)
$$
by starting with the root(s) at the highest level and working downwards.
Using analogous results for $\bar \cJ_{-\a}^L$, we find
$$
\del_{-\a}V=0\quad(\a\in\Ph^+\bs\D) \qquad \del_{-\a}V=-\n_{-\a}V
 \quad (\a\in\D).
\eqno(4.3)
$$
Hence we have, from Eqs. (3.20), (3.23), (and corresponding results for
$\a\rightarrow-\a$)
$$\eqalignno{
X_{\a}^RV=X_{\a}^LV=0, \quad(\a\in\Ph^+\bs\D)\qquad
&X_{\a}^RV=X_{\a}^LV=\m_{\a}V, \quad(\a\in\D)\cr
X_{-\a}^RV=X_{-\a}^LV=0, \quad(\a\in\Ph^+\bs\D)\qquad
&X_{-\a}^RV=X_{-\a}^LV=\n_{-\a}V,\quad(\a\in\D).&(4.4)\cr}
$$
The physical state condition Eq. (2.17) becomes, using Eqs. (3.15) and
(4.4)
$$
\bigl(({\del\over{\del R}},{\del\over{\del R}})
+2({1\over{\l}}\d+\l\r,{\del\over{\del R}})
-2(\l^2-{1\over k})\sum_{\a\in\D}M_{\a}e^{-{1\over{\l}}(R,H_{\a})}-1\bigr)V
=0 \eqno(4.5)
$$
where $M_{\a}$ is given by Eq. (2.10), $\d$ is given in Eq. (2.12),
and we define $\l$,
$R$, ${\del\over{\del R}}$ and $\r$ by
$$
\l={1\over{\sqrt{k-h}}},\quad
R={1\over{\l}} r^iH_i, \quad {\del\over{\del R}}=\l H^i{\del\over{\del r^i}},
\quad  \r={1\over2}\sum_{\a\in\Ph^+}H_{\a},\eqno(4.6)
$$
with $H^i$ as given in Eq. (3.6).
Following Ref.\dvv, we now wish to interpret the differential operator in Eq.
(4.5) as the Hamiltonian operator derived from an action
$$
S(r^i)={1\over{8\p}}\int d^2x\sqrt\eta
\bigl(\eta^{\m\n}\del_{\m}r^i\del_{\n}r^jg_{ij}
+V(r^i)+D(r^i)R^{(2)}\bigr) \eqno(4.7)
$$
where $R^{(2)}$ denotes the two-dimensional Ricci scalar.
The conformal invariance conditions for the tachyon potential $V(r^i)$ in
 Eq. (4.7) are
equivalent to the field equations derived from the string effective action,
which is given to lowest order by
$$
S(V)=\int dr e^{-2D(r)}\sqrt g (g^{ij}\del_iV\del_jV-2V^2+\ldots)
 \eqno(4.8)
$$
where the ellipsis represents a polynomial in $V$ which is determined by
non-perturbative effects\ref\das{S. R. Das and B. Sathiapalan, \prl56 (1986)
2664; A. A. Tseytlin, \plb264 (1991) 311.}
\ref\arkIII{A. A. Tseytlin, \plb241 (1990) 233.}. However, these
non-perturbative effects can be ignored for the purposes of reproducing
results derived in conformal theory, which are predicated upon
a renormalisation scheme in which non-perturbative effects are
absent\ref\arkIV{A. A. Tseytlin, private communication.}. There are
also higher-order perturbative corrections to $g_{ij}$ in Eq. (4.8), which
can be deduced from the tachyon $\b$-functions calculated in Ref.\ref\jjm
{I. Jack, D. R. T. Jones and D. A. Ross., \npb307 (1988) 531; I. Jack, D. R.
 T. Jones and N. Mohammedi, \npb332 (1990) 333}. However, these
corrections can be absorbed in a field redefinition of the metric and dilaton
\jjp.
Identifying the physical state condition Eq. (4.5) with the $\s$-model
conformal invariance conditions, we deduce that the derivative parts of the
operator in Eq. (4.5) should be given by
$$
{-1\over{e^{-2D}\sqrt{g}}}\del_ie^{-2D}\sqrt{g}g^{ij}\del_j. \eqno(4.9)
$$
 Comparing Eqs. (4.5), (4.9), we see that
$$\eqalignno{
\del_{\m}r^i\del_{\n}r^jg_{ij}&=(\del_{\m}R,\del_{\n}R) \cr
D(r^i)&=({1\over{\l}}\d+\l\r,R).&(4.10)\cr}
$$
However, it is not clear how to derive the correct quantum form of the
potential term $V(r^i)$ itself from a comparison of the field equations for
Eq. (4.8) with Eq. (4.5). Previous calculations\ref\dhoker{J. Goldstone
(unpublished); E. D'Hoker and R. Jackiw, \prd26 (1982) 3517}
 indicate that the
original exponential interactions in Eq. (2.9) are modified by multiplicative
renormalisations of $M_{\a}$ arising from one-loop perturbative effects.
Other authors\ref\gris{M. T. Grisaru, A. Lerda, S. Penati and
D. Zanon, \plb234 (1990) 88; \npb342 (1990) 564; \npb346 (1990) 264.}
 have found non-perturbative renormalisations for exponential interactions
 with imaginary exponents,
but these do not arise in the case of exponential interactions with real
exponents such as we consider here\arkIII.
We can now reconstruct the fully conformally invariant action for the
 Toda field theory as
$$
S(R)={1\over{8\p}}\int d^2x\sqrt{\eta}
\bigl(\eta^{\m\n}(-\del_{\m}R,\del_{\n}R)
+({1\over{\l}}\d+\l\r,R)R^{(2)}+E(R)\bigr). \eqno(4.11)
$$
where $E(R)$ denotes the interaction potential.
Using the well-known formula for the central charge for the Coulomb gas
representation, we find the central charge for the Toda field theory to be
 given by\dubI
$$
c=\rank\cG+12({1\over{\l}}\d+\l\r,{1\over{\l}}\d+\l\r). \eqno(4.12)
$$
For the case of a simply-laced Lie algebra (with the roots
normalised by $<\a,\a>=2$), for which $\d=\r$, we have, upon using the
Freudenthal-de Vries strange formula
$$
c=\rank\cG+({1\over{\l}}+\l)^2h{\rm dim}\cG, \eqno(4.13)
$$
which is the correct central charge for the Toda field theory
for a simply-laced Lie algebra at the quantum
level\man.

Finally, we note that the canonical expression for the Toda action may be
achieved by setting
$$
R=\sum_{\a\in\D}\wr^{\a}H_{\a}  \eqno(4.14)
$$
in Eq. (4.11), and using Eqns. (A.5), (A.7)-(A.9). We find
$$\eqalignno{
S(\wr)&={1\over{8\p}}\int d^2x\sqrt{\eta}\bigl( -
\sum_{\a,\b\in\D}\eta^{\m\n}
{A_{\a\b}\over{<\a,\a>}}\del_{\m}\wr^{\a}\del_{\n}\wr^{\b}\cr
&\quad+\sum_{\a\in\D}({1\over{\l<\a,\a>}}+{\l\over2})\wr^{\a}R^{(2)}
+E(\wr)\bigr).
&(4.15)\cr}
$$
\newsec{Conclusions}

We have obtained the correct central charge at the quantum level for the Toda
field theory, for both the simply-laced\man\ and non-simply-laced\dubI\ cases,
in what we believe is a succinct and elegant fashion. We note that the
dilaton and central charge for the Toda theory were calculated in
Ref. \arkIII\
by assuming an action of the general Toda form and solving the conformal
invariance condition for the tachyon (i. e. the field equation for Eq. (4.8)).
Our approach is somewhat analogous since again the tachyon conformal
 invariance condition plays a central r\^ole. However the main feature of our
method is that it shows how the correct quantum properties of the Toda
theory can be derived from the WZW model.
As we mentioned in the previous section, it is not clear if one can
determine the quantum-corrected exponential interactions $E(R)$ from the
non-derivative part of the operator in Eq. (4.5).
 This deserves further investigation.

In any case, Eq. (4.5) may have an interesting direct physical
interpretation. The case of the Liouville model was discussed in Ref.\dvv.
There it was pointed out that Eq. (4.5) was equivalent to a
mini-superspace Wheeler-DeWitt equation for a loop amplitude, interpreted
as a wave function for two-dimensional quantum gravity\ref\moore{G. Moore,
N. Seiberg and M. Staudacher, \npb362 (1992) 365.}. In the same way as
the Liouville model describes the effective induced action for
 two-dimensional quantum gravity in the conformal gauge,
 it appears that Toda field theory describes
the effective induced action for $W$-gravity in the conformal gauge\ref\goer
{J. de Boer and J. Goeree, \plb274 (1992) 289; \npb381 (1992) 329.},
 and so one may speculate that
Eq. (4.5) can be interpreted as the mini-superspace Wheeler-DeWitt equation
for a wave-function in $W$-gravity.

The Gauss decomposition Eq. (2.3) corresponds to the canonical grading of the
Lie algebra (in which elements of the form $g_0$ in Eq. (2.4) generate an
abelian subgroup of $\cG$). One may also consider non-canonical gradations
of the Lie algebra which lead to a modified gauss decomposition for which
$g_0$ generates a non-abelian subgroup. Upon going through the process of
reducing the WZW model by imposing constraints
\dubI\dubIII,
 this yields the so-called
non-abelian Toda field theory\natI. At the classical level, the non-abelian
toda
theory based on the Lie algebra $B_2$ has a metric sector which reproduces
Witten's string black hole solution\gervais\sbh. This solution is only
 classically
conformally invariant; the exact solution, conformally invariant to all orders,
was constructed in Ref.\dvv\ using the methods we have used here. It would be
interesting to apply these methods to the $B_2$ non-abelian Toda field theory
to check whether the exact conformally invariant version of this theory
agrees in the metric sector with the exact string black hole solution of
Ref.\dvv. Moreover, the central charge for a general non-abelian toda theory
was calculated in Refs. \dubIII, \hay. The methods we have used here should
also provide an efficient means of reproducing these results.
\bigskip\bigskip
\centerline{{\bf Acknowledgements}}
We thank A. A. Tseytlin,
D. R. T. Jones and H. Osborn for very useful discussions. This work
 was supported by the SERC.
\appendix{A}{Lie Algebra conventions}
We introduce a basis $\{H_i, i=1,\ldots \rank\cG\}$ for the Cartan subalgebra
of $\cG$, and also step operators $\{E_{\a}, \a\in\Ph\}$ corresponding to the
 roots $\a$. We then have
$$
[H_i,H_j]=0, \qquad [H_i,E_{\a}]=\a(H_i)E_{\a} \eqno(A.1)
$$
We define the Cartan-Killing form by
$$
(X,Y)=\tr(\ad X\ad Y)      \eqno(A.2)
$$
where $X$, $Y$ are elements of the Lie algebra and $\ad$ is defined in
Eq. (3.17).
The Cartan-Killing form is then defined by
$$
\g_{ij}=(H_i,H_j),\quad \g_{\a\b}=(E_\a,E_\b) \eqno(A.3)
$$
We then define $H_{\a}$ by
$$
(H_{\a},H)=\a(H)  \eqno(A.4)
$$
for all $H$ in the Cartan sub-algebra. This enables one to define an inner
product on the roots by
$$
<\a,\b>=(H_{\a},H_{\b})      \eqno(A.5)
$$
and we also find
$$
[E_{\a},E_{-\a}]=\g_{\a(-\a)}H_{\a} \eqno(A.6)
$$
The Cartan matrix $A_{\a\b}$ is then defined by
$$
A_{\a\b}=2{<\a,\b>\over{<\b,\b>}}.  \eqno(A.7)
$$
It can be shown that
$$
\r=\sum_{\a,\b\in\D}A^{\a\b}H_{\b} \eqno(A.8)
$$
and
$$
\d=2\sum_{\a,\b\in\D}{A^{\a\b}\over{<\a,\a>}}H_{\b}. \eqno(A.9)
$$
We also make use of the Freudenthal-de Vries strange formula
$$
24(\r,\r)=h<\psi,\psi>{\rm dim}\cG     \eqno(A.10)
$$
where $\psi$ is the highest root and $h$ is the dual Coxeter number.

\listrefs
\end